\documentclass[letter,prl,twocolumn]{revtex4-1}
\usepackage{fullpage}
\usepackage{draftcopy}
\usepackage{graphicx}
\usepackage{verbatim}
\usepackage{color}
\usepackage{amsmath}
\usepackage{algpseudocode}

\newcommand{\beq}{\begin{eqnarray}}
\newcommand{\eeq}{\end{eqnarray}}
\newcommand{\be}{\begin{equation}}
\newcommand{\ee}{\end{equation}}

\begin{document}

\title{Molecular-Atomic Transition in the Deuterium Hugoniot with Coupled Electron Ion Monte Carlo}
\author{Norm M. Tubman$^1$, Elisa Liberatore$^2$, Carlo Pierleoni$^3$, Markus Holzmann $^4$, David M. Ceperley$^1$}

\affiliation{$^1$Department of Physics, University of Illinois, Urbana, Illinois 61801, USA\\
$^2$ EPFL, Lausanne, Switzerland\\
$^3$ Department of Physical and Chemical Sciences, University
of L'Aquila and CNISM UdR L'Aquila, Via Vetoio 10, I-67010
L'Aquila, Italy\\
 $^4$  LPTMC, UMR 7600 of CNRS, Universit Pierre et Marie Curie, 75005 Paris, France\\}

\date{\today}
\begin{abstract}
We  have performed accurate simulations of the Deuterium
Hugoniot using Coupled Electron Ion Monte Carlo (CEIMC). Using
highly accurate quantum Monte Carlo
methods for the electrons,
we study the region of maximum compression along the principal Hugoniot, where the system undergoes a continuous transition from a molecular fluid to a monatomic fluid.
We include all relevant physical corrections so that a direct
comparison to experiment can be made.   Around 50 GPa we found a maximum compression of 4.85, roughly 10\% larger than previous theoretical predictions and experimental data but still compatible with the latter because of their large uncertainty.
\end{abstract}
\maketitle

\newpage

The study of high pressure hydrogen is particularly interesting
as progress in the field has come about from difficult
experiments under extreme conditions and computationally
expensive quantum simulations~\cite{rmp}.
Experiments on hydrogen under high pressure have direct
implications for planetary science: laboratory setups attempt
to recreate the extreme conditions which describe planetary
formation and equilibrium properties of planetary
interiors~\cite{Stevenson82,Militzer08,Nettelmann08,Guillot05,Fortney04_2}.
Improvements from both theory and experiment
have been essential to creating our current understanding of
the hydrogen phase diagram~\cite{Vorberger2007,PNAS2010,rev1,rev2}.
A key experimental technique to probe hydrogen under extreme
conditions is dynamic compression via shock wave generation.  The principal Hugoniot~\cite{hugo1,hugo2,rev1,rev2} is
determined by shocking a material from an initial state to a
state of higher pressure, temperature, and density. The locus
of points reachable in such an experiment is determined by
conservation laws and initial conditions with few theoretical
assumptions. Shock experiments often use
deuterium instead of hydrogen as a means to reach higher 
overall pressures~\cite{knudson2004,boriskov1,nellis2,hicks1,grishechkin1,belov1,wang1,vanthiel1,knudson1,knudson2009}.  Theoretical methods used so far to investigate this interesting region of phase diagram are based on density functional theory (DFT) which is expected to describe molecular dissociation and metallization with only limited accuracy.

 In this work we present highly accurate
quantum Monte Carlo (QMC) results for the crossover between the molecular liquid to
monatomic liquid along the principal deuterium Hugoniot.  We find that the maximum compression through this molecular dissociation crossover is 10\% larger than previous predictions from DFT.   
Among the many computational methods used in high pressure
electronic structure simulations, QMC is
considered among the highest quality~\cite{grossman1,tubman1,tubman2,ericneu},
with Fixed-Node Projector Quantum Monte Carlo being the most
accurate~\cite{foulkes1,baroni1,luke1,clay2014,krogel1,huihuo1,krogel2,bart1}. The Coupled Electron Ion Monte
Carlo method (CEIMC)~\cite{dewing1,dewing2,pierleoni1}  uses QMC to determine the electronic ground state energy, with  a finite temperature
sampling of the nuclear coordinates on the resulting Born-Oppenheimer energy surface~\cite{born1,tubman-born,adam3}.  
This can be combined with other techniques such as DFT and variational Monte Carlo (VMC) for wave function generation, path integral Monte Carlo (PIMC) for the ions~\cite{ceperley1,ceperley2}, correlated
sampling for calculating energy differences~\cite{lester1}, and
Reptation Quantum Monte Carlo~ (RQMC) for calculating unbiased
estimators~\cite{baroni1}, all of which are included in our
simulations.

\textit{Deuterium Hugoniot:} Shock experiments are used to
determine the equation of state of a material that is in
an initial state at a known energy, pressure and volume:
($E_{0}$,$P_{0}$,$v_{0}$). The zeros of the Hugoniot function
$H(v,T)$ determines the final conditions $E$,$P$,$v$ as \be
H(v,T)=e(v,T)-e_0 + \frac{1}{2} (v-v_0) (P(v,T)+P_0) =0 \ee where $v$
is the atomic volume, $e(v,T)$ is the internal energy/atom and $P(v,T)$ is
the pressure. We use initial conditions (0.167 g/cm$^{3}$,
22K, 1.24$\times 10^{-4}$GPa) matching some of the previous experiments~\cite{knudson2004}. In
atomic units this corresponds to $v_0= 135.15 a_0^3,$ $P_0=4.2
\times 10^{-9}$ a.u., $ e_0=-0.583725$ Ha/atom. and $r_s=3.18353$.

\begin{table}[ht]
\caption{ CEIMC-RQMC estimates of the principal Hugoniot: Pressures,
r$_{s}$, deuterium mass density, compression and temperature }
\begin{tabular}{ccccc}
\hline
 P(GPa) &r$_{s}$&$\rho_{d}$(gcm$^{-3})$&$\rho_{d}/\rho_{0}$&T(10$^{3}K)$ \\
 \hline
 18(1) &  2.019(5) &0.654(5) & 3.91(3) &  4    \\
 32(1) &  1.909(9) &0.773(9) &4.63(6) &  6    \\
 39(1) &  1.882(3) &0.807(4) &4.83(2) &  8    \\
 48(1) &  1.880(3) &0.810(4) &4.85(2) &  10    \\
 66(1) &  1.895(1) &0.791(2) &4.73(1) &  15    \\
 \end{tabular}
\label{tab:ceimc}
\end{table}

\begin{figure}[tbp]
\centering
\includegraphics[scale=0.4]{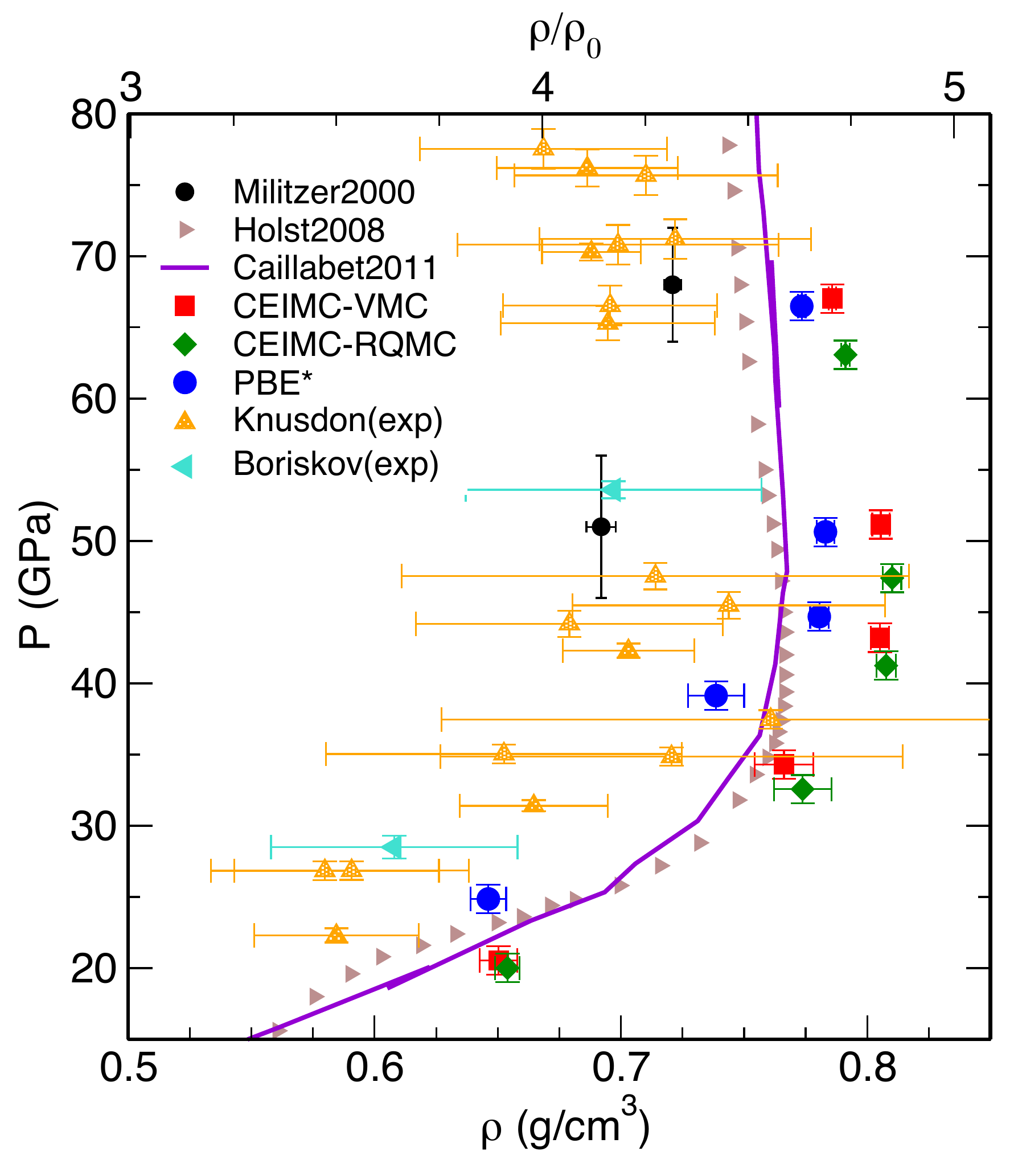}
\caption{The principal Deuterium Hugoniot compared to previous
theoretical  and experimental studies. Holst~\cite{holst1} and
Caillabet~\cite{caillabet1} are PBE-DFT simulations,
Militzer~\cite{militzer1} is a PIMC simulation. The PBE* results
are generated by solving Hugoniot equation using
the VMC-CEIMC configurations  but computing energy and pressure
with PBE-DFT. Knudson~\cite{knudson2009} and Boriskov~\cite{boriskov1} are experimental results. The initial density for the Boriskov experiment is slightly higher, 0.171 g/cm$^{3}$, than the other Hugoniots. \label{fig:press1}}
\end{figure}

\begin{figure}[]
\centering
\includegraphics[scale=0.4]{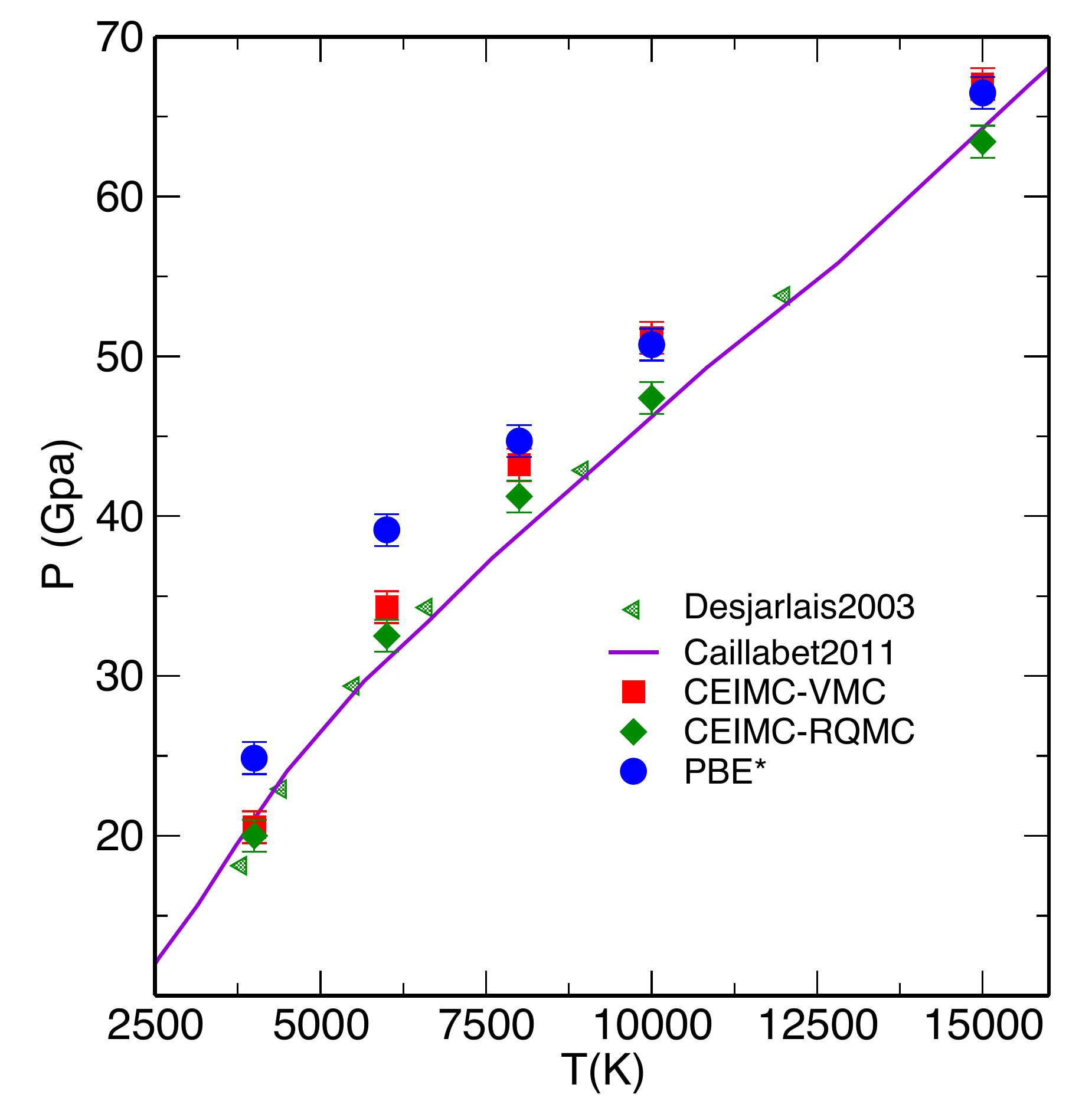}
\caption{Pressure vs temperature along the Deuterium
Hugoniot compared to previous theoretical studies.
\label{fig:temp1}}
\end{figure}
To calculate the Hugoniot in the region of interest, we perform
simulation at fixed densities in the range $ 1.80 \leq r_s \leq 2.00$ and along isotherms in the range
$4,000K \leq T \leq 15,000K$.  Fitting $H(v,T)$ at fixed T to a quadratic
polynomial of $v$ we solve for $H=0$.  Our results are shown in
Figure \ref{fig:press1}, Figure \ref{fig:temp1} and Table
\ref{tab:ceimc}. The main result is the CEIMC-RQMC curve.  The CEIMC-VMC and PBE* predictions are also results from this work and will be discussed later.  

 Previous theoretical results have  been generated from a
variety of different methods which include DFT, PIMC, and wave
packet
MD~\cite{desjarlais1,holst1,caillabet1,militzer1,Filinov01,Lenosky00,Bonev04_1,khair1,knaup1}.
So far DFT Hugoniots have been performed with the PBE
functional; two of them are shown in the Hugoniot plots (Holst2008~\cite{holst1}, and Caillabet2011~\cite{caillabet1}). The DFT studies generally show similar behavior, but other methods that involve different approximations generally do not agree with these results, especially in the crossover region.  For instance, previous PIMC calculations~\cite{militzer1} with variational density matrix
nodes show significantly different behavior in this region of phase space. 
The present calculation is unique in that we are able to control many of the relevant errors with a
wave function based approach.

There are notable differences between previous DFT predictions
and our results, especially when considering the maximum
compressibility region between 
8,000K and 10,000K. The DFT Hugoniots consistently show a maximum
compression of 4.4,
but both the VMC and RQMC results show a maximum compression of
$\sim$4.85.  It might be expected that most DFT functionals would
struggle to capture the physics of this crossover, as the
energies of bond breaking of just two hydrogen atoms are poorly
described with many density functionals such as PBE.
However, despite the differences between the QMC and DFT results, it is not clear that this is the origin  
of the discrepancy. The average distance between hydrogen atoms at these
pressures is much smaller than that of traditional bond breaking
physics and it is not clear how important issues of  self-interaction and symmetry-breaking effects are for this transition.

There has been an extensive amount of experimental work in measuring the Hugoniot for deuterium~\cite{knudson2004,boriskov1,nellis2,hicks1,grishechkin1,belov1,wang1,vanthiel1,knudson1,knudson2009} and hydrogen~\cite{nellis1,weir1,Holmes95,Nellis83,Sano2010}. 
Experimental results from Knudson et al. ~\cite{knudson2009} and Boriskov et al.~\cite{boriskov1} are plotted for comparison in Figure \ref{fig:press1}.  
  
\begin{figure}[tbp]
\centering
\includegraphics[trim=1.5cm 0cm 1.9cm 0cm,clip=true,scale=0.45]{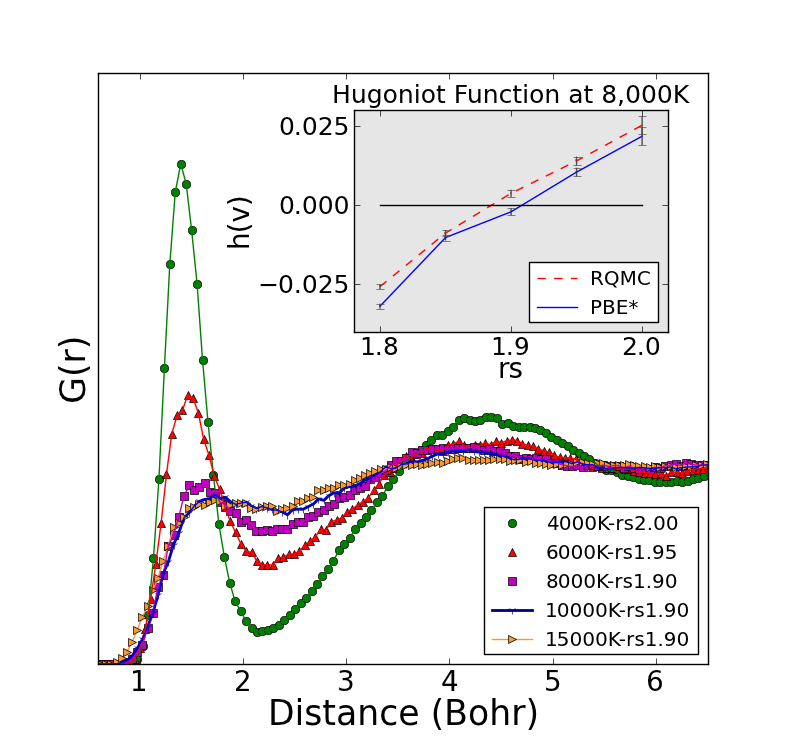}
\caption{Proton-Proton pair correlation functions near the Hugoniot points.  The large peak in the first pair correlation plot represents a fully paired molecular phase.  This peak disappears continuously as a transition to a monatomic liquid is observed along the Hugoniot.  Inset:  The Hugoniot function plotted at 8,000K, for RQMC and PBE.  The VMC Hugoniot function is not plotted as it lies nearly on top of the RQMC.\label{fig:press}}
\end{figure}

Evidence of the bond-breaking crossover is given in Fig.
\ref{fig:press} where we present the proton-proton correlation
function along the Hugoniot.
The minimum temperature at which we observe the breaking of hydrogen molecules is density dependent. 
At the highest density in this work $r_{s} =1.80$, a small increase in the temperature
over 4,000K causes a transition to the monatomic phase whereas
in the lowest density systems $r_{s} = 2.00$, the crossover
does not occur until the system is above 10,000 K.  
  
To establish the accuracy of our results, we now discuss the details of our simulations.  It is crucial that the main part of our simulations, the QMC electronic structure calculations, are performed with high quality wave functions.  The wave function form we use consists of a single Slater determinant for each spin component and a correlation part consisting of one, two, and three body Jastrows. We use DFT Kohn-Sham orbitals from quantum espresso package~\cite{QE-2009}, for each configuration of the ions, in the Slater determinants. 
Analytical expressions from RPA for both correlation and backflow functions are employed~\cite{holzmann2,dewing1,pierleoni2,alder1}, which are complemented by empirical expressions, with a few variational parameters, that preserve the short and large distance limits~\cite{dewing1,pierleoni2}.  To mitigate the computational effort we optimize the wave function parameters over an ensemble of statistically independent configurations at a fixed density and at thermal equilibrium.  We find that using this form of the wave function yields energies within 1mHa/atom of a full optimization of each configuration individually.

\begin{table}[ht]
\caption{Differences (mHa per atom) between RQMC
energies and VMC energies at various densities and temperatures. The "Avg Err" is the mean absolute error (MAE) of the energies over configurations, and "Rel Err" is the MAE between configurations after the energies have been shifted by the average energy difference of the entire set.   $r_s$ is the Wigner-Seitz radius in bohr
radii. Temperatures are expressed in Kelvin.  Configurations are sampled with VMC-CEIMC and the RQMC energy differences are calculated on a set of 100 configurations.}
\begin{tabular}{|cc|cccccc|}
 \hline
 Avg Err& $r_s$ &4000(K) &6000 & 8000 &10000 &15000\\
\hline
 &1.8 &  2.7(3) &  3.2(3) &  3.4(3) &  3.6(3) &  4.1(3) &  \\
 &1.85 &  3.0(3) &  3.2(3) &  3.6(3) &  3.8(3) &  4.5(3) &  \\
 &1.9 &  3.2(3) &  3.5(3) &  3.9(3) &  4.5(3) &  4.6(3) &  \\
\hline
 Rel Err & $r_s$ &4000(K) &6000 & 8000 &10000 &15000\\
\hline
 &1.8 &  0.16(1) & 0.21(2) & 0.27(2) & 0.34(3) & 0.42(3) &  \\
 &1.85 &  0.18(2) & 0.22(2) & 0.19(2) & 0.32(2) &0.48(3) & \\
 &1.9 &  0.16(2) & 0.24(3) & 0.28(5) & 0.33(5) & 0.44(4) & \\

 \hline
\end{tabular}
\label{tab:1}
\end{table}

We perform simulation of systems of 54 protons and 54 electrons at fixed volume and temperature. CEIMC runs are performed with energy differences from VMC. To demonstrate the quality of our wave function we select a number of statistical independent configurations 
generated during the CEIMC run and compare VMC against RQMC energies, as shown in Table~\ref{tab:1}.

In order to calculate an accurate Hugoniot the errors in the energy and pressure need to be consistent across
densities for a given temperature.  This consistency is apparent in our data for all the temperatures considered in this work.   
Just as important, the largest discrepancy is less than 1~mHa/atom, which is small enough as to not influence the results beyond the final error bars on our calculated Hugoniot curves.

Additionally in Table~\ref{tab:1} we report  relative energy errors. 
The relative energies between configurations is important as it determines whether we are sampling an accurate thermal distribution for the ions.  Table \ref{tab:1} shows that these relative energy differences are significantly less than 1 mHa/atom.  To test the quality of our sampled distribution, we used reweighting~\cite{reweigh} at 8000K for rs=1.85,1.90 over 1500 nuclear configurations getting an efficiency of ~0.5, high enough to testify the significant overlap of RQMC-generated and VMC-generated thermal distributions of the ions.

\textit{Approximations and corrections:}
Due to the nature of these simulations, there are other issues of accuracy beyond the electronic structure.  In this section we discuss how we correct for the finite size of the
simulation cell, thermal effects of the electrons, nuclear quantum effects, and corrections coming from RQMC.

The corrections for using a finite simulation cell
can be divided into single particle and many particle effects. Single
particle finite size effects can be accounted for by using twisted 
boundary conditions~\cite{lin1}.  We used a fixed grid of ($4\times4\times4$) twisted angles. The many body finite size
effects can be estimated~\cite{chiesa1,drummond1} by extrapolating the
small wave length limit of the charge-charge structure factor $S_{qq}(k)$.
The corrections comprises a kinetic energy contribution,
$\Delta K = 3/\sqrt{16~r_{s}^{3}}$, and a potential
energy contribution 
$\Delta V =3~r_{s}^{-3} lim_{k\rightarrow 0}\left[S_{qq}(k)/k^2\right]$.
Corrections to the pressure can also be introduced~\cite{martin1}, $\Delta P = \left[(2~\Delta K + \Delta
V)\rho\right]/3$.  

Corrections from electronic thermal effects are computed with PBE-DFT.  
  Given a set of representative configurations, we have run DFT with a smearing of the electronic density over an ensemble of single particle orbitals weighted by the Fermi-Dirac distribution. 
Once the energies and pressures have been calculated in DFT, there are two types
of corrections that can be made. The thermal DFT energy and pressure corrections 
 can be added in directly to correct our QMC energies and pressures.   The second correction involves reweighting the configurations with the electronic entropy term, $TS$, so as to incorporate the effects of using Mermin finite temperature DFT~\cite{mermin1965}.    
  It is not clear in whether including these thermal corrections improve our estimates, as both the DFT band gaps and pressures are important in determining thermal effects.  The problems with DFT band gaps is well studied~\cite{martin2,louie1,louie2,foulkes2}, and in the next section we show that the DFT pressure errors are significant.      
We have included the first correction by calculating the thermal effects on 50 configurations for each temperature/density considered in this work.  We also tested re-weighting with the electronic entropy at 8000K and observed no effect within our error bars.

Nuclear quantum effects (NQE) can be explicitly taken
into account with PIMC techniques.
However because the temperatures considered are rather large and such calculations are computationally more expensive, we spot-checked with PIMC simulations at only two densities ($r_{s}
= 1.80, 2.00$) and at T=8,000K. We found no
effect on the energies and pressures within our error bars. 
Further we have estimated NQE at T=4,000K by the molecular zero point energy $\hbar\omega_0/2$ with $\omega_0$ fitted to the observed bond distribution~\cite{holst1}. Corrections to the energy and pressure are significant at this lower temperature but the global effect on the Hugoniot is within our present error bars. 

Lastly, as previously mentioned, we perform VMC-CEIMC
calculations and add in the RQMC energy and pressure as a correction. RQMC
calculations of the energy and pressure are extrapolated to infinite
projection time ($\beta$) and zero time step ($\Delta\tau$).

\begin{table}[ht]
\caption{ Pressure MAE (GPa) between RQMC  and
VMC/PBE .  Configurations are sampled with VMC-CEIMC
and the PBE and RQMC pressure differences are calculated on a
subset of the generated configurations.}
\begin{tabular}{|cc|cc|ccc|}
\hline
 &$r_s$ &8K-VMC& 10K-VMC&8K-PBE &10K-PBE\\
\hline
  &1.8 &  2.8(2) &  2.8(2) &  6.4(2) &  6.3(2) &  \\
 &1.85 &  2.3(2) &  2.7(2) &  3.1(2) &  3.7(2) &  \\
 &1.9 &  2.7(2) &  3.6(2) &  6.1(3) &  5.7(3) &  \\
\hline

 \end{tabular}
\label{tab:2}
\end{table}

\begin{table}[ht]
\caption{ Energy MAE (mHa per atom) between RQMC and
VMC/PBE.  Configurations are sampled with VMC-CEIMC
and the PBE and RQMC energy differences are calculated on a
subset of the generated configurations.}
\begin{tabular}{|cc|cc|ccc|}
\hline
 &$r_s$ &8K-VMC& 10K-VMC&8K-PBE &10K-PBE\\
\hline
 &1.8 &   3.4(3) &  3.6(3) &  4.0(3) &   3.7(3) &  \\
 &1.85 &  3.6(3) &  3.8(3) &  3.6(3) &  4.2(3)&  \\
 &1.9 &  3.9(3)&  4.5(3)  &  4.2(3)&  3.5(3)&  \\
\hline

 \end{tabular}
\label{tab:3}
\end{table}

 A question that remains to be clarified is the origin of the
differences between the PBE functional and QMC.  The simulations described in this work involve a main QMC calculation of the energy and pressure before corrections are taken into account.  We can replace the QMC calculation with PBE-DFT, as a test of how the functional behaves differently from our QMC results, while the finite size corrections and the thermal corrections remain fixed.  Specifically we took a
subset of our VMC-CEIMC configurations and used PBE-DFT to
recalculate all the energies and pressures.  
All electron PBE calculations were performed with a sufficiently high plane wave cutoff (500 Ry) to
converge the energies and pressures.   
We perform the PBE calculations at zero temperature and with the same k-point sampling
 that we used for our QMC twist averaging.  
With this data we recalculated the Hugoniot. The results are shown in the
Figures \ref{fig:press1} and \ref{fig:temp1} as PBE*.  We are most interested
in the temperatures at 8,000 K and 10,000 K where our CEIMC calculations exhibit the largest compression
The PBE* curve at both these temperatures are less compressed (4.6), than our VMC/RQMC results.
  We can understand this result by considering the energy and pressure errors in Table
\ref{tab:2} and Table \ref{tab:3} for our PBE and VMC.  The VMC and PBE energy errors are actually quite close, and consistently agree within error bars for this part of the phase diagram.  
A trace of the energies for the individual configurations suggests
that the two methods may generate very similar ensembles of
ionic configurations.  The change in the PBE* curve mainly comes from errors in the
pressure as shown in Table \ref{tab:2}. These pressure errors  are more than twice as large as the VMC and their magnitude fluctuates  significantly at different densities. This is in comparison to the VMC pressure errors which are not only smaller, but are consistent with the energy errors in such a way that the VMC and RQMC Hugoniot functions are very similar.  
  A comparison of PBE* and the VMC/RQMC Hugoniot functions are plotted in the inset of Figure \ref{fig:press}.

\textit{Discussion/Conclusions:}
In this work we have performed a calculation of the principal
Hugoniot of deuterium in the region of the crossover between
the molecular to atomic phase.  
 Our results show that deuterium is more compressible
than estimated on the basis of previous PBE-DFT simulations.  We suggest
that a large part of the difference arises from errors in the
DFT pressures, and our results suggest that both energy and pressure
errors become more significant at temperatures below 8000K.
This represents one of the first works for dense hydrogen in which all
the relevant physical effects were taken into account without the possibility
for any large uncontrolled errors.   In fact, algorithmic and computational
advances in the last several years have been significant and we
are in a position to improve upon many previous predictions in
high pressure physics.  

\textit{Acknowledgments}: We thank L. Schulenburger, L. Benedict, W. Nellis, E. Brown, T. Mattsson, H. Changlani,  I. Kylanpaa, and M. Knudson for useful discussions. This work was supported by DOE DE-NA0001789. NA0001789. CP was supported by the Italian Institute of Technology (IIT) under the SEED project Grant 259 SIMBEDD. Computer time was provided by XSEDE, supported by the National Science Foundation Grant No.~OCI-1053575, and by PRACE projects 2011050781 and 2013091918.

 \bibliography{refs}{}

\end{document}